\documentclass[10pt, journal, compsoc]{IEEEtran}

\usepackage{cite}
\usepackage{mathrsfs}
\usepackage{graphicx}
\usepackage{textcomp}
\usepackage{hyperref}
\usepackage[ruled,linesnumbered]{algorithm2e}
\usepackage{amsmath,amssymb,amsthm, amsfonts}

\begin{document}

\title{Comprehensive Survey towards Security Authentication Methods for Satellite Communication Systems}

\author{
Yunfei Meng,~\IEEEmembership{Member,~IEEE,}
Changbo Ke,~\IEEEmembership{Member,~IEEE,}
Zhiqiu Huang,~\IEEEmembership{Member,~IEEE,}
\thanks{
\newline This work was supported by Doctoral Fellow Foundation of Qingdao Binhai University (Grant No.BS2022A10), partially supported by General Program of National Natural Science Foundation of China (Grant No.62072253) and General Program of Natural Science Foundation of Jiangsu Province (Grant No.BK20221327).
\newline Yunfei Meng is with the College of Information Engineering, Qingdao Binhai University, 266555 Qingdao, China (e-mail: 06354@qdbhu.edu.cn).
\newline Changbo Ke is with the School of Computer Science and Technology, Nanjing University of Posts and Telecommunications, 210023 Nanjing, China (brobo.ke@njupt.edu.cn).
\newline Zhiqiu Huang is with the College of Computer Science and Technology, Nanjing University of Aeronautics and Astronautics, 210016 Nanjing, China (zqhuang@nuaa.edu.cn).
}
\thanks{ }}



\maketitle

\begin{abstract}
Satellite communication systems (SatCom) is a brand-new network that uses artificial Earth satellites as relay stations to provide communication services such as broadband Internet access to various users on land, sea, air and in space. It features wide coverage, relatively high transmission rates and strong anti-interference capabilities. Security authentication is of crucial significance for the stable operation and widespread application of satellite communication systems. It can effectively prevent unauthorized access, ensuring that only users and devices that pass security authentication can access the satellite network. It also ensures the confidentiality, integrity, and availability of data during transmission and storage, preventing data from being stolen, tampered with, or damaged. By means of literature research and comparative analysis, this paper carries out on a comprehensive survey towards the security authentication methods used by SatCom. This paper first summarizes the existing SatCom authentication methods as five categories, namely, those based on cryptography, Blockchain, satellite orbital information, the AKA protocol and physical hardware respectively. Subsequently, a comprehensive comparative analysis is carried out on the above-mentioned five categories of security authentication methods from four dimensions, i.e., security, implementation difficulty and cost, applicable scenarios and real-time performance, and the final comparison results are following obtained. Finally, prospects are made for several important future research directions of security authentication methods for SatCom, laying a well foundation for further carrying on the related research works.
\end{abstract}


\section{Introduction}
\IEEEPARstart{S}{atellite} communication system (SatCom) refers to a system that uses artificial Earth satellites as relay stations to achieve communication between different locations on Earth. It forwards radio signals through satellites to achieve long-distance and wide-area communication coverage, and is widely used in fields such as broadcasting, television, telephony, the Internet, and military communications\cite{maral2020satellite}. Currently, due to their characteristics of wide coverage and being free from geographical limitations, Satellite communication systems are widely applied in various fields such as radio and television broadcasting, Internet access, maritime communication, aviation communication, emergency rescue, the Internet of Things, distance education and medical care, scientific research, finance and commerce, navigation and positioning, as well as energy management.

\subsection{SatCom Architecture}
The basic architecture of a satellite communication system mainly consists of three parts: the space segment, the ground segment, and the user segment. These three parts work together to achieve the global communication coverage and data transmission functions of the satellite communication system\cite{tedeschi2022satellite}.

\textbf{(1) The space segment} is the core component of the satellite communication system, mainly composed of communication satellite constellations\cite{zhang2024satellite}. These satellites are distributed at different orbital altitudes, including Low Earth Orbit (LEO), Medium Earth Orbit (MEO), and Geostationary Earth Orbit (GEO), etc. Satellites at different orbital altitudes have their own characteristics and advantages. Low-orbit satellites have the advantages of low transmission delay, low link loss, and flexible launch. They can achieve high-speed data transmission and are very suitable for the development of satellite communication system services. For example, SpaceX's Starlink project uses a large number of low-orbit satellites to provide high-speed Internet access services to global users. Medium-orbit satellites strike a good balance between coverage and transmission performance and can provide a relatively wide area coverage\cite{xu20226g}. Geostationary Earth Orbit satellites\cite{zhang2023survey}, due to their geostationary position relative to the Earth, can achieve continuous coverage of specific areas and are often used in fields such as radio and television signal transmission and fixed communication. Satellites are interconnected through inter-satellite links, forming a huge space-based network\cite{zhang2021satellite}. Inter-satellite links can use microwave or laser communication technology. Laser communication has the advantages of high transmission rate, large communication capacity, strong anti-electromagnetic interference performance, and high confidentiality. It can effectively improve the data transmission efficiency and security between satellites and is an important direction for the future development of inter-satellite links\cite{yuan2024global}.

\textbf{(2) The ground segment} is the connection hub between the satellite communication system and the ground communication system, mainly including the satellite tracking and control network, authentication servers, network control centers, etc. The satellite tracking and control network is responsible for the real-time monitoring and control of the satellite's orbit, attitude, working status, etc., to ensure the normal operation of the satellite\cite{peng2025space}. The gateway station is the interface between the satellite network and the ground communication system. It is responsible for forwarding the data transmitted by the satellite to the ground communication system and sending the data of the ground communication system to the satellite\cite{peng2025space}. The gateway station is connected to the Internet backbone network or other ground communication systems through ground communication lines to achieve the interconnection and intercommunication between the satellite communication system and the ground network. The network control center is responsible for the management and scheduling of the entire satellite communication system system, including satellite resource allocation, user access management, network performance monitoring, etc\cite{huang2024future}.

\textbf{(3) The user segment} is the terminal part of the satellite communication system, including various types of user terminal devices, such as fixed terminals, mobile terminals, and portable terminals. These terminal devices communicate with the satellites in the space segment through satellite signals to achieve user access to the satellite communication system. Fixed terminals are usually used for Internet access in fixed places such as homes and enterprises. For example, rural families can achieve high-speed and stable Internet connection by installing satellite fixed terminals and carry out e-commerce, online education, and other businesses. Mobile terminals are mainly applied to moving carriers such as vehicles, ships, and aircraft to provide real-time communication services for mobile users. For example, during the voyage of an ocean-going ship, it can maintain communication with the land through a satellite mobile terminal to obtain weather information, navigation data, etc. Portable terminals are convenient for users to use outdoors or during movement, such as satellite phones and portable satellite Internet access devices. They play an important role in scenarios such as wilderness exploration and emergency rescue\cite{peng2025zero}.

\subsection{Unique Characteristics of SatCom}
Satellite communication systems have a series of unique characteristics, making them play an important role in the communication field.

\textbf{(1) Satellite communication systems have the remarkable characteristic of wide coverage.} Traditional ground communication systems, such as fiber-optic networks and mobile communication base stations, are limited by geographical conditions and infrastructure construction and are difficult to cover remote areas, oceans, deserts, and other regions. However, satellite communication systems are not restricted by these factors\cite{lakew2023review}. By deploying satellites in Earth's orbit, they can achieve seamless global coverage. Whether in remote mountainous areas, vast oceans, or boundless deserts, as long as within the satellite signal coverage area, users can access the satellite communication system to obtain information and services. This characteristic makes satellite communication systems irreplaceable in providing communication services to remote areas, supporting offshore operations, and ensuring aviation communication. For example, in some island countries, due to the scattered geographical environment, the construction of ground communication systems is costly and difficult, and satellite communication systems have become an important means for these countries to achieve national communication coverage\cite{zhou2022research}.

\textbf{(2) The transmission rate of satellite communication systems is relatively fast}. With the continuous development of satellite communication technology, especially the application of High Throughput Satellites (HTS)\cite{tedeschi2022satellite}, the transmission rate of satellite communication systems has been greatly improved. Some advanced satellite systems can provide transmission rates of up to several Gbps or even higher, meeting the needs of users for high-rate services such as high-definition video and large-data transmission. In the aviation field, satellite communication systems provide high-speed network connections for aircraft, allowing passengers to smoothly watch high-definition videos and conduct video conferences during the flight. This makes satellite communication systems highly competitive in meeting users' needs for high-speed data transmission and can provide users with a better communication experience\cite{shen2024integrated}.

\textbf{(3) Satellite communication systems have strong anti-interference capabilities}. Satellite communication uses microwave or laser communication technology, and the signals are transmitted in the space outside the atmosphere, being less affected by ground electromagnetic interference\cite{ahmad2022security}. In contrast, ground communication systems are easily affected by electromagnetic interference, terrain, and other factors, resulting in a decline in signal quality or even interruption. In some areas with complex electromagnetic environments, such as military bases and industrial parks, satellite communication systems can provide more stable and reliable communication services. In the event of natural disasters, ground communication systems may be severely damaged. Due to their independence from ground infrastructure, satellite communication systems can maintain communication during critical moments and provide important communication support for emergency rescue, disaster monitoring, etc. For example, in the event of Earthquakes, floods, and other disasters, satellite communication systems can quickly establish temporary communication systems to ensure communication between the rescue command center and the disaster area, providing strong support for the smooth progress of rescue work\cite{kang2024survey}.

\subsection{Security Threats Faced by SatCom}
However, due to the deep integration of satellite communication systems and ground networks, ground network attacks have become the main security threat faced by satellite communication systems. Due to the broadcast characteristics of the satellite's uplink and downlink, malicious users can eavesdrop to steal the personal identification information transmitted by legitimate users and then use identity spoofing attacks to illegally access the satellite communication system. Once a malicious user achieves illegal access, they may launch destructive flooding attacks such as DDoS or TCP SYN against the satellite communication system\cite{ahlawat2024key}\cite{tirmizi2022hybrid}. Due to its own payload limitations, satellite base stations usually have limited computing power and it is impossible to deploy complex traffic analysis and security defense mechanisms\cite{manulis2021cyber}. Therefore, after being attacked, it is extremely easy to cause a significant decline in network Quality of Service (QoS) or even trigger a single-point collapse. Illegal access may also lead to the leakage of sensitive information, causing huge losses to users and related institutions. In the military field, if the enemy illegally accesses the satellite communication system and obtains military confidential information, it may pose a serious threat to national security
\cite{chen2022security}. In the commercial field, if a company's trade secrets are illegally obtained, it may put the company at a disadvantage in market competition\cite{kumar2024review}.

This paper is structured as follows. Section 2 outlines the importance of security authentication for SatCom and the entire process. Section 3 reviews the existing SatCom security authentication methods according to five types, namely, those based on cryptography, Blockchain, satellite orbital information, the AKA protocol and physical hardware. Section 4 comprehensively compares the five types of methods mentioned in Section 3 in terms of four dimensions, namely security, implementation difficulty and cost, applicable scenarios, and real-time performance, and the final comparison results are following obtained. Section 5 prospects several important future research directions in the research field of SatCom security authentication. Finally, Section 6 concludes the entire paper.

\section{Overview of Security Authentication for SatCom}

\subsection{Importance of Security Authentication for SatCom}
Security authentication, as a crucial means of ensuring the security of satellite communication systems, is of vital significance for the stable operation and widespread application of satellite communication systems\cite{suhaimi2024state}. It can effectively prevent unauthorized access. Only users and devices that have passed security authentication can access the satellite network, thereby preventing unauthorized users from obtaining network resources and sensitive information. Security authentication can safeguard information security, ensuring the confidentiality, integrity, and availability of data during transmission and storage, and preventing data from being stolen, tampered with, or damaged\cite{yang2025communications}. In the military field, satellite communication systems are tasked with transmitting important military information. Security authentication can prevent enemy forces from stealing military secrets and safeguard national military security. In the commercial field, financial institutions use satellite communication systems for data transmission\cite{chen2023research}. Security authentication can ensure the security of financial transactions and protect users' property security. Security authentication also helps to enhance user trust. When users know that their operations and data on the satellite communication system are protected by security authentication, they will have more trust and reliance on satellite communication system services, thus promoting the expansion and popularization of satellite communication system services\cite{ren2024secure}.

\subsection{Security Authentication Process for SatCom}
The security authentication process of the satellite communication system can be described as shown in Figure 1. Specifically:
(1) A legitimate user first needs to register personal and terminal identification information (ID) on the authentication server. The authentication server assigns a private key (SK) and a public key (PK) to the user, and saves the identification information ID and PK.
(2) The user encrypts the identification information ID using SK to generate a data packet AU = SK(ID) for security authentication, and then connects to the satellite in space through the satellite modem.
(3) After receiving the AU data packet, the communication satellite is not responsible for security authentication. It only relays the AU data packet to other satellites and finally forwards it to the ground gateway (Gateway).
(4) The ground gateway receives the AU data packet and forwards it to the authentication server for analysis.
(5) The authentication server decrypts the AU data packet using the user's public key PK to obtain the user identification information ID, and then compares it with the user registration ID information saved in the server. If they match, the security authentication is passed, and a session key (K) for encrypted communication is generated. Then, the user's public key PK and the session key K are distributed to the ground gateway.
(6) After receiving (PK, K), the ground gateway dynamically assigns an IP address to the user terminal through the built-in DHCP server or static IP address management function. Then, it modifies its own firewall forwarding rules to ensure that the user terminal can access the target network. It then encrypts (K, IP) using the user's public key PK to generate an encrypted data packet AG = PK(K, IP) and forwards it to the user terminal through the satellite.
(7) After receiving the AG data packet, the user terminal decrypts the AG data packet using SK to obtain (K, IP), and then uses the obtained session key K and the IP address to establish an encrypted communication link with the ground security gateway.
(8) When the current communication session ends, the session key K will automatically become invalid. The user needs to connect to the authentication server again for security authentication to obtain a new session key and IP address.
\begin{figure*}[htbp]
\centering
\scalebox{0.25}{\includegraphics{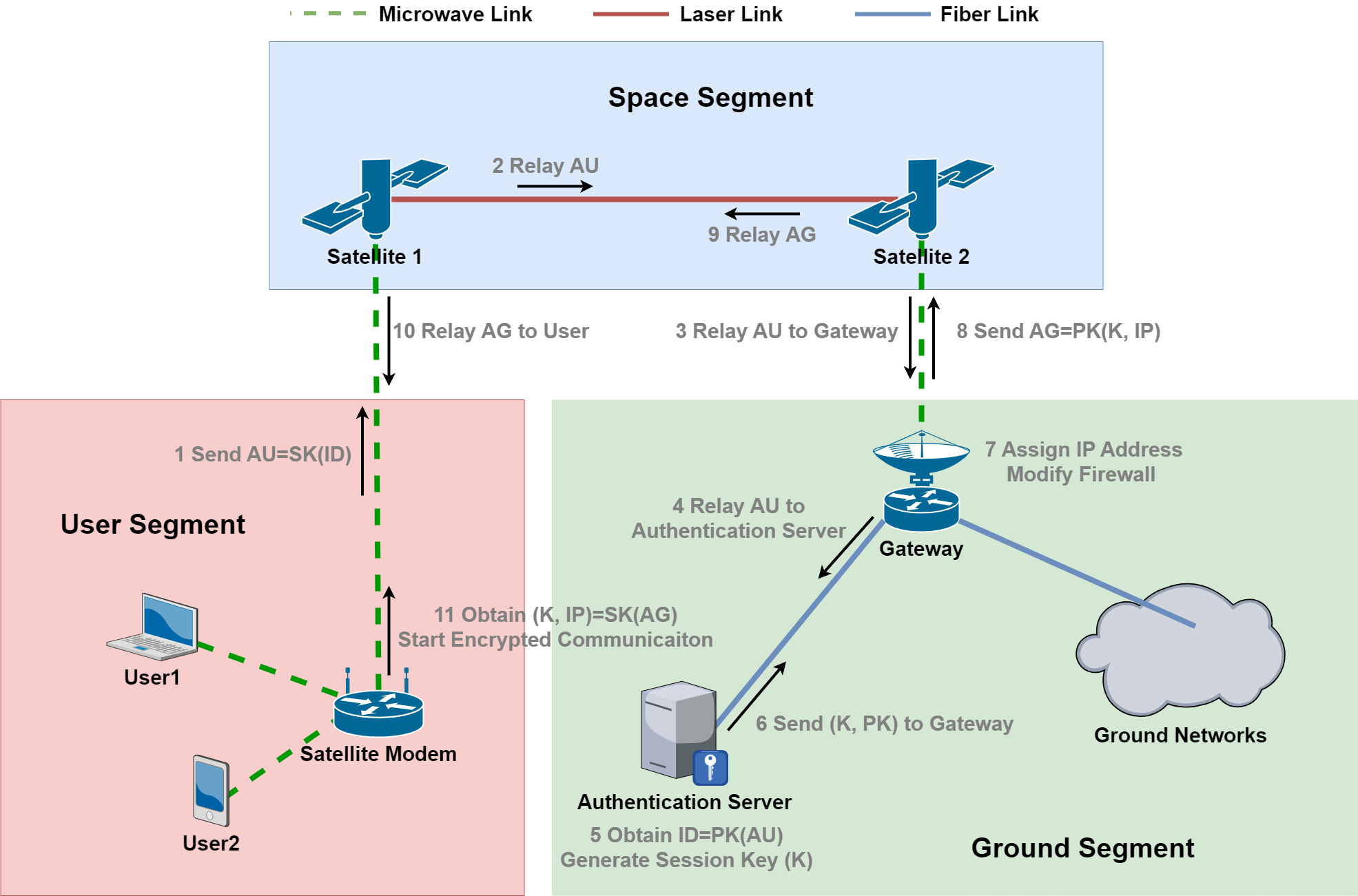}}
\caption{Security Authentication Process for SatCom.}
\end{figure*}

\section{Research Status of Security Authentication Methods for SatCom}

By investigating large amount of relevant proposals, the security authentication methods used by satellite communication systems could be summarized as five categories, i.e., cryptography based methods, Blockchain based methods, satellite orbit information based methods, AKA protocol based methods as well as physical hardware based methods.

\subsection{Cryptography Based Methods}
The basic principle of using cryptography to conduct security authentication for satellite communication system terminals is to utilize the efficiency and security of cryptography to achieve terminal identity verification and communication encryption. The authentication process of this type of method is as follows: The terminal generates a public-private key pair using cryptography (such as dual elliptic curves, etc.), signs the message with the private key, and sends an authentication request to the authentication server. The authentication server stores the public keys and unique identifiers of all legitimate terminals and uses the terminal's public key to verify the signature. If they match, the security authentication is passed. After successful authentication, the terminal conducts encrypted communication with the server via satellite.

Altaf et al.\cite{altaf2021novel} proposed an authentication protocol based on elliptic curve cryptography (ECC), which consists of two main stages: registration and login authentication. In the registration stage, the user registers with the Network Control Center (NCC), and the NCC generates a smart card for the user. During the login authentication, the user sends an authentication request to the NCC through the smart card. After the NCC verifies the user's identity, both parties generate a session key by exchanging random numbers. The verification methods adopted are informal and formal security analyses.
Liu et al.\cite{liu2022secure} proposed an authentication protocol based on elliptic curve cryptography, which includes stages such as initialization, registration, group key negotiation, initial authentication, and handover authentication. In the initialization stage, the NCC and UE generate global public-private key pairs respectively. In the registration stage, the AP, AP group manager, and UE register with the Network Control Center (NCC) respectively and obtain corresponding keys and pseudo-IDs. In the group key negotiation stage, APs in the same group request the GM to allocate group keys and authentication certificates. In the initial authentication stage, the UE and AP conduct identity verification and session key negotiation by exchanging messages. In the handover authentication stage, the UE conducts fast handover authentication with the target AP based on the previously obtained group information and credentials.The verification methods adopted are to use the eCK model\cite{eCK_model} to prove the security of the session key in the authenticated and unauthenticated link models.
Li et al.\cite{li2022lightweight} proposed a lightweight handover authentication mechanism based on elliptic curve cryptography, which includes four stages: system setup, registration and key distribution, handover authentication, and private key update. In the system setup stage, the AMF generates system parameters. In the registration stage, mobile users and access nodes register with the core network and obtain public-private key pairs. In the handover authentication stage, mobile users and access nodes conduct two-way identity verification and session key negotiation by exchanging messages, supporting batch authentication. In the private key update stage, the AMF updates the system key and user private key through the public channel. The verification method adopted is to use the AVISPA tool\cite{AVISPA} for formal security verification to prove that the protocol meets the security objectives.
S. Li et al.\cite{li2024enabling } proposed a secure networking authentication scheme based on elliptic curve public key cryptography and Chebyshev polynomials, which is divided into three stages: node registration stage, SAT-HAP networking authentication stage, and SAT-UAV networking authentication stage, achieving efficient mutual authentication and session key negotiation between nodes. The verification methods adopted are informal security analysis and formal security simulation using Scyther tool\cite{Scyther} to verify the security of the scheme.
Ali Khan etc.\cite{khan2022raks} proposed a robust authentication and key agreement scheme based on elliptic curve cryptography (ECC) and hash functions. When the user registers, the NCC generates a smart card for the user, which contains relevant encrypted information. In the login and authentication stage, the user enters the identity, password, and biometric features. The smart card calculates and verifies the relevant information. After passing the verification, it conducts key negotiation with the NCC to establish a shared session key. The verification methods adopted are formal security analysis under the random oracle model (ROM)\cite{ROMeprint} to prove the security of the scheme. The AVISPA tool is used to simulate and verify man-in- the-middle attacks and replay attacks.
Yulei Chen, Jianhua Chen\cite{chen2021enhanced} proposed an enhanced dynamic authentication scheme based on ECC and hash functions to address the weaknesses of Altaf et al.'s scheme\cite{altaf2021novel}. When the user registers, the NCC checks the user's identity and stores the relevant information in the smart card. In the login and authentication stage, the user inputs the identity, password, and biometric features. After the smart card passes the verification, it conducts authentication and key negotiation with the NCC. The verification methods adopted are formal security proof under the random oracle model (ROM) based on the BPR2000 adversary model\cite{BPR2000} to prove the security of the scheme.
Tao et al.\cite{tao2023demand} proposed an on-demand anonymous access and roaming authentication protocol. Ordinary nodes use a short group signature algorithm based on bilinear pairing to achieve unlinkable authentication; low-energy-consuming nodes use a lightweight batch authentication protocol to achieve fast authentication and can resist denial-of-service (DoS) attacks by malicious nodes; an efficient cross-domain roaming authentication protocol is designed to reduce authentication delay. The verification methods adopted are to use the ProVerif tool\cite{ProVerif} for formal verification, simulate the operation of the protocol, and verify the security of the protocol against known active and passive attacks.
Kumar et al.\cite{kumar2022note} improved the scheme proposed by Chen\cite{chen2021enhanced}. During the authentication process, by modifying the relevant calculation and verification steps, the problem that users cannot authenticate with the Network Control Center (NCC) and establish a session key in the original scheme is solved. The verification methods adopted are informal security analysis to analyze the security of the improved scheme.
Soltani et al.\cite{soltani2022robust} proposed an authentication and session key negotiation protocol based on elliptic curve cryptography (ECC). Through multiple rounds of message interaction between the user and the NCC, two-way authentication and secure negotiation of the session key are achieved. The verification methods adopted are to use the Automated Validation of Internet Security Protocols and Applications (AVISPA) tool for formal verification.
Guo et al.\cite{guoj2023srakn} proposed a secure roaming authentication and key negotiation protocol (SRAKN) based on ECC, which supports conditional anonymity and batch verification. Through multiple stages such as system setup, user registration, pre-negotiation, roaming authentication, batch verification, and password- biometric update, secure authentication and key negotiation are achieved among roaming users, satellite nodes, and foreign ground control stations (FTCS). The verification methods adopted are to prove the semantic security of the negotiated session key under the random oracle model (ROM) and the AVISPA tool.

Yu Wang etc.\cite{wang2021lightweight} proposed a lightweight and secure authentication protocol suitable for the railway space-ground integrated network, which is an improvement and extension based on the standard authentication protocol EPS-AKA. In the space network initialization and registration stage, relevant key information is generated using the elliptic curve cryptosystem (ECC) and registration is completed. The initial access authentication protocol uses a public key encryption algorithm to solve the problem of plain text transmission of important information. The handover authentication protocol realizes fast handover authentication with the help of the Chinese Remainder Theorem. The verification methods prove through security analysis.
Junyan Guo etc.\cite{guoj2021provably} proposed a provably secure access and handover authentication protocol based on ECC. In the system setup stage, system public parameters are generated. In the registration stage, public-private key pairs are generated for legitimate nodes. In the mutual authentication stage, mutual authentication between the user and the ground station is achieved through an intermediate satellite node, and the legitimacy of the satellite relay node is confirmed. In the handover stage, solutions are designed for two scenarios: satellite handover and ground station handover. The ground station handover supports multi-user batch handover authentication. The verification methods adopted are formal security analysis using the random oracle model (ROM) and the AVISPA tool.
Junyan Guo etc.\cite{guo2021secure} proposed a three-factor anonymous roaming authentication protocol based on ECC, which supports user authentication when the user is in the local domain and when roaming to a foreign domain. In the system setup stage, an elliptic curve and a base point are selected, and public-private key pairs are generated. In the registration stage, the user and the satellite submit registration requests to the local TCS. In the login stage, the user inserts the smart card and enters relevant information for verification. In the mutual authentication stage, it is carried out in two cases. Through multiple rounds of message interaction, mutual authentication and session key negotiation are achieved. The verification methods adopted are formal security analysis using BAN Logic\cite{BAN_original}, the ROR model\cite{BellareRogaway1993}, and the AVISPA tool.
Huang et al.\cite{huang2020mutual} proposed the encryption-based mutual authentication and key update (EMAKU) protocol, which includes mutual authentication between satellites and the Ground Control Center (GCC), mutual authentication between satellites, and the key update process for LEO satellites. In mutual authentication, symmetric keys and message authentication codes are used for identity verification. When updating keys, GEO satellites are used as a bridge to update the keys of LEO satellites. The verification method adopted is to conduct a security analysis taking replay attacks and man-in-the-middle attacks as examples to demonstrate the security of the protocol.
Yang et al. \cite{yang2023lkaka}proposed a lightweight authentication and key agreement protocol for satellite-to-satellite (S2S) communication based on satellite location keys and symmetric crypto systems. Location keys are constructed using satellite orbital parameters and location information, and combined with long-term shared keys to achieve two-way authentication and key agreement between satellites. Two location key update schemes are designed to deal with situations such as satellite link disconnection. The verification method adopted is to use informal security analysis and the Scyther tool to prove the security of the scheme.
Wu Zhijun et al.\cite{zhijun2024bdsec } proposed the BDSec authentication protocol. The protocol uses the Chinese commercial cryptography SM series of algorithms. The protocol consists of two sub-protocols: authentication initialization and information authentication, involving processes such as BeiDou satellite initialization requests and responses, receiving user requests for authorized identities, user key pair requests and feedback, and the embedding, broadcasting, and receiving verification of authentication information. In the authentication initialization sub-protocol, each entity completes identity authentication and parameter allocation. The verification method adopted is to use SVO Logic\cite{SVO_journal} reasoning to analyze the security of the protocol and prove that it can meet the security requirements.
Tian Minqiu et al. \cite{tian2024trust}proposed a satellite Internet of Things terminal authentication method based on trust assessment. For the two situations of re-authenticating a strongly authenticated terminal within a short period and being trusted by a strongly authenticated node, a dynamically adjusted trust metric and evaluation model are described according to direct trust and indirect trust. Using hash chains and SM2 digital signature technology, an access authentication protocol based on trust assessment is designed, and an appropriate authentication mechanism is adopted according to the device trust level. The verification method adopted is to comprehensively analyze the security of the protocol through non-formal methods and the Tamarin tool\cite{Tamarin}.
Hehzad Ashraf Chaudhry et al.\cite{chaudhry2021lightweight} proposed a lightweight authentication protocol (LSMP-MTS) applicable to 6G-IoT-enabled maritime transport systems. The protocol includes stages such as initialization, HAPS registration, vessel registration, and mutual authentication, and uses lightweight primitives of symmetric encryption and hash operations. The verification method adopted is to verify its security through the random oracle model (ROM) and informal analysis.
Chen et al. \cite{chen2009self} proposed a self-verifying authentication mechanism, which consists of three phases: initialization, registration, and authentication. In the initialization phase, the Network Control Center (NCC) selects the cryptographic system parameters based on the discrete logarithm problem. In the registration phase, the NCC generates a signature for the user and derives the user's master key, and stores the relevant information in the user's smart card. In the authentication phase, the user calculates the Message Authentication Code (MAC) and sends it to the NCC. The NCC verifies the user's identity by verifying the MAC, derives the session key, and generates a new temporary identity to send to the user.
Yuanyuan Yang et al.\cite{yang2023fhap} proposed a Fast Handover Authentication Protocol (FHAP) for high-speed mobile terminals in 5G satellite-terrestrial integrated networks. User terminals form temporary groups within the same carriage and are managed by a Relay Node (RN). Based on the Chinese Remainder Theorem (CRT), the 5G core network configures pre-authentication information for multiple access nodes to achieve fast handover authentication between the user terminal and the Target Access Node (TAN). The verification methods adopted are to use Scyther and BAN Logic tools for security analysis.
Wang et al.\cite{wang2023security} proposed a secure and efficient authentication key negotiation scheme for space-ground integrated railway networks based on Software-Defined Networking (SDN). In the initial authentication phase, a lightweight two-way authentication mechanism based on the Number Theory Research Unit (NTRU) is adopted to ensure the legal access of the On-Board Unit (OBU) to the network. Hash chains and the Chinese Remainder Theorem (CRT) are used to design key pre-generation and fast distribution mechanisms respectively to reduce the key transmission burden. In heterogeneous networks, a Hash-based Message Authentication Code is used to achieve unified handover authentication. The verification methods adopted are to BAN Logic proof and informal security analysis to prove the scheme.

\subsection{Blockchain Based Methods}
The basic principle of using Blockchain to authenticate satellite communication system terminals is to leverage the decentralized, tamper-proof, and transparent characteristics of the Blockchain or Smart Contracts to verify the uniqueness and legitimacy of terminal identities. The authentication process of such methods is as follows: The terminal generates a public-private key pair and registers the public key and identification information on the Blockchain\cite{Bitcoin_whitepaper} or Smart Contract\cite{Ethereum_yellowpaper}. The terminal uses the private key and identification information to generate a digital signature, sends an authentication request to the authentication server, and attaches the signature information. The authentication server queries the terminal's public key and identification information by accessing the Blockchain. The server decrypts using the public key and verifies the terminal's signature and identity information. If they match, the security authentication is passed. After successful authentication, the terminal conducts encrypted communication with the server via satellite.
Deng et al. \cite{deng2021Blockchain} proposed an authentication protocol (BAPC) consisting of three stages: registration, initial authentication, and handover authentication. In the registration stage, users purchase cryptocurrency from the authentication center, and the authentication center stores the transaction records on the Blockchain. During initial authentication, the satellite confirms the user's identity based on the user's transaction information on the Blockchain. In the handover authentication stage, the satellite determines whether to continue providing services to the user by verifying the transactions on the Blockchain. The verification method adopted is to conduct experiments on the NS2 network simulation platform\cite{NS2_website}, comparing the handover authentication with other protocols.
Wang et al.\cite{wang2022ntru} proposed an authentication scheme based on Blockchain and certificateless encryption (BC-Authen). They used a consortium Blockchain to record registration activities and utilized certificateless encryption to calculate key pairs for smart devices. The authentication process includes system initialization, registration, access authentication, and fast access authentication stages. Smart devices authenticate their identities through the public keys registered on the Blockchain. The verification method adopted is to conduct a security analysis from the aspects of authentication security and conditional privacy.
C. Li et al.\cite{li2024Blockchain} constructed an architecture composed of satellites and ground devices, integrating the authentication and privacy protection structure of the communication system. Information communication, registration, authentication, and revocation are achieved in stages. The Blockchain is used to record key parameters, and an asymmetric encryption algorithm is employed to transmit keys, enhancing data transmission security. The verification method adopted is to evaluate the performance of the architecture through simulation experiments.
Xiang Liu, Anjia Yang et al.\cite{liu2022Blockchain} proposed a decentralized anonymous authentication scheme based on Smart Contracts and cryptographic tools. Users conduct anonymous authentication by presenting secret knowledge bound to randomized verifiable credentials. The authentication task is carried out by satellites, reducing the interaction with the ground network control center (NCC) and thus reducing the authentication delay. When users register, multiple NCCs generate partial credentials for them. After collecting at least t partial credentials, users aggregate them into a final identity credential. During the authentication and prepayment stages, users randomize the credentials and calculate relevant parameters. Satellites confirm the legitimacy of users by verifying zero-knowledge proofs and pairing operations. The verification method adopted is to use zero-knowledge proof (ZKP)\cite{ZKP_original} technology to ensure that users prove they possess secret knowledge without revealing other information.
Ting Xiong et al.\cite{xiong2022Blockchain} adopted a Blockchain-based identity management mechanism, designing two types of identities for satellites: permanent and temporary. Permanent identities are used for intra-constellation communication, while temporary identities are used for inter-constellation collaboration to protect satellite privacy. Satellites register their temporary identities with the local identity management and Blockchain server (IMBS). After Blockchain consensus, they use the temporary identities for authentication during collaboration. The verification method adopted is to conduct a formal security analysis using BAN Logic to prove the scheme.
Torky et al.\cite{torky2022Blockchain} proposed the Proof of Space Transactions (PoST) protocol based on Blockchain, modeling space transactions as Space Digital Tokens (SDT). The validity of SDT is verified through the Blockchain to achieve authentication of communication between satellite constellations. The verification method adopted is to implement a prototype on the Ethereum and evaluate the performance of the protocol.
Li et al.\cite{li2021research } constructed an authentication and privacy protection scheme for satellite communication systems based on Blockchain technology. The Blockchain is used to record the communication and control instructions between devices and manage permissions. Combined with an off-chain database, data and permissions are separated to ensure data security. Satellites forward the collected information to ground base stations, which record the key parameters on the distributed Blockchain and revoke the certificates of malicious nodes. The verification method adopted is to conduct a simulation experiment.
Jianfeng Guan et al.\cite{guanj2021Blockchain} proposed a Blockchain-assisted secure and lightweight authentication (BSLA) scheme. The scheme is divided into five stages: initialization, registration, broadcast, unicast, and handover. It adopts a Blockchain-based distributed architecture, dispersing the private key generation power to multiple Blockchain nodes to solve the private key escrow problem in the traditional identity-based cryptosystem (IBC). In the registration stage, users obtain partial private keys through the Blockchain and combine them to generate the final private key. The authentication process is based on identity signatures to achieve mutual authentication between mobile nodes (MN) and proxy authentication centers (PAC). The verification method adopted is to analyze the security of the protocol.
Ting Xiong et al.\cite{xiong2022delay} proposed a delay-aware cooperative caching scheme for on-chain authentication in Low Earth Orbit (LEO) satellite communication systems. LEO satellites are divided into multiple clusters, and satellites within each cluster cooperate to cache the Blockchain, using the consistent hashing algorithm for caching. By analyzing query delay and synchronization delay, the clustering problem is modeled as a coalition formation game, and a distributed delay-aware coalition formation algorithm (DAC) is designed to find the optimal coalition division. The verification method adopted is to compare indicators through simulation experiments.
Zhijun Wu et al.\cite{wu2023Blockchain } proposed an authentication scheme for Global Navigation Satellite System (GNSS) civil navigation messages based on a consortium Blockchain, using domestic cryptographic algorithms. An anti-spoofing architecture including a ground network layer, a user network layer, a Blockchain network layer, and a Blockchain is constructed. The authentication model is based on identity cryptography (IBC) and public key infrastructure (PKI), achieving the authentication of information sources and information content through the interaction and certificate verification between nodes. The authentication protocol includes steps such as users sending authentication requests to BCDA and the information interaction and verification between nodes. The verification method adopted is to prove the security of the protocol through SVO Logic.
Yihan Lei et al.\cite{han2021handover } proposed a handover authentication technology based on distributed consensus and Ticket for space information networks. Based on the Raft consensus algorithm, the distributed management of user trust states is carried out, and the trust authorizations are shared and stored on the chain to achieve the traceability of user handover behaviors. A Ticket model including a holder retention part and a submission part is designed to assist in authentication. A handover authentication scheme combining Raft consensus and Ticket is proposed. Based on the management of user trust states, fast handover authentication is achieved through Ticket verification. The verification method adopted is to conduct a security analysis from three aspects: Raft consensus, Ticket, and the authentication process to prove the security of the scheme.
Y. Ma et al.\cite{ma2021Blockchain } constructed a secure communication framework for satellite network systems using Blockchain and Smart Contracts. The Blockchain provides the same tamper-proof and traceable distributed ledger for all nodes in the network, realizing distributed identity authentication. Node registration modules, message communication modules, and integrity check modules are designed, and Smart Contracts are used to regularly verify the integrity of key data of nodes. The verification method adopted is to implement a private Blockchain on a Linux server, deploy the satellite secure communication mechanism based on Blockchain and Smart Contract, and use the Truffle framework and Ethereum JavaScript Tests RPC for testing.
Li Xueqing et al.\cite{ li2024Blockchain } proposed an inter-satellite networking authentication technology based on Blockchain, including Blockchain-based satellite identity authentication and inter-satellite message authentication. Satellite temporary identities are registered on the Blockchain, and satellite identity authentication is completed by comparing the information on the chain. The elliptic curve digital signature algorithm is used to sign and verify inter-satellite messages. The verification method adopted is to demonstrate the security of the identity authentication method through informal security analysis.
Dongxiao Liu et al. \cite{liu2022Blockchain}proposed a collaborative credential management scheme for anonymous authentication in Space-Air-Ground Integrated Vehicular Networks (SAGVN) based on Blockchain (SAG-BC). A Blockchain-based distributed system setup (DSS) and a collaborative credential issuance (CCI) scheme are designed. Zero-knowledge proofs and concise on-chain commitments are used to achieve efficient verifiability and incentive mechanisms. The verification method adopted is to conduct experiments on a real consortium Blockchain to demonstrate the feasibility and efficiency of the scheme and analyze its security.

\subsection{Satellite Orbit Information Based Methods}
The basic principle of securely authenticating satellite communication system terminals based on satellite orbit information is to use satellite orbit parameters (such as position, velocity, time, etc.) as authentication factors and combine them with cryptographic techniques to ensure the legality of terminals and the security of communication. The authentication process of such methods is as follows: First, the satellite regularly broadcasts its own orbit information; the terminal generates a public-private key pair, obtains the satellite orbit information by receiving satellite signals, performs a hash calculation on the orbit information, signs the hash value with the private key, and sends an authentication request to the authentication server. The authentication server stores the public keys and unique identifiers of all legitimate terminals, verifies the signature using the terminal's public key, calculates the hash value of the orbit information and compares it. If they are consistent, the security authentication is passed. After successful authentication, the terminal conducts encrypted communication with the server via the satellite.
Jedermann et al. \cite{jedermann2021orbit} used satellite orbit information and TDOA (Time Difference of Arrival) signature for authentication. Multiple ground receivers collect the time differences of satellite signal arrivals to form TDOA signatures, which are compared with the expected signatures calculated from the satellite orbit information. If they match, the authentication is successful. The verification method adopted is to evaluate the authentication performance through simulation experiments, and use ANOVA tool\cite{R_ANOVA} to study the impacts of different factors.
Gautam et al.\cite{gautam2024robust} proposed an inter-satellite authentication protocol based on elliptic curve cryptography and satellite orbital coordinates, which includes three stages: system initialization, registration, and authentication. In the authentication stage, satellites generate authentication messages using their own position information and shared keys, verify each other's identities, and negotiate session keys. The verification methods adopted are informal security analysis and formal security proof using the ROR model and BAN Logic.
Thakur et al.\cite{thakur2024efficient} proposed a security authentication protocol based on the elliptic curve discrete logarithm problem. They generated position keys using satellite orbital coordinates and combined elliptic curve cryptography to achieve mutual authentication and key negotiation between satellites. The verification methods adopted include formal proof and informal analysis.
Abdrabou et al.\cite{abdrabou2022authentication} used Doppler shift (DS) and received power (RP) characteristics for satellite authentication, and adopted hypothesis testing (threshold method) or machine learning (OCC-SVM) methods to distinguish between legitimate and illegal satellites. During the communication process, the DS and RP values are continuously updated to provide reliable authentication. The verification method adopted is to use real satellite data obtained from the STK toolkit\cite{AGI_STK} for performance evaluation.
Ozan Alp Topal et al.\cite{topal2022physical} proposed a physical layer authentication (PLA) method based on Doppler shift (DS) in the literature "Physical Layer Authentication for LEO Satellite Constellations" for the authentication of communication links between low-Earth orbit (LEO) satellites. The Doppler shift of the satellite is used as a digital fingerprint, and the receiving satellite verifies the identity of the transmitter by comparing the measured Doppler frequency with the reference Doppler frequency of the legitimate transmitter. Multiple satellites fuse decisions to determine the final authentication result. The verification method adopted is to evaluate the authentication performance by deriving analytical expressions.
Mohammed Abdrabou et al.\cite{abdrabou2022physical} proposed an adaptive physical layer authentication scheme based on machine learning in the literature "Physical Layer Authentication for Satellite Communication Systems Using Machine Learning". Doppler shift (DS) and received power (RP) are used as features, and one-class classification support vector machine (OCC-SVM) is combined to authenticate low-Earth orbit satellites. Continuous authentication of satellites is achieved by continuously updating the training data. The verification method adopted is to calculate the missed detection rate (MDR), false alarm rate (FAR), and authentication rate (AR) using the confusion matrix to evaluate the performance of the scheme.
Mohammed Abdrabou, T. Aaron Gulliver et al.\cite{abdrabou2022threshold } proposed a threshold-based physical layer authentication scheme for low-Earth orbit (LEO) satellite authentication in the literature "Threshold-Based Physical Layer Authentication for Space Information Networks". Doppler frequency spread (DS) and received power (RP) are used as features, and hypothesis testing is used to distinguish between legitimate and illegal satellites. In the system model, the ground mobile satellite communication (OTMSC) station first conducts upper-layer authentication, and then uses the DS and RP values for physical layer authentication. During the authentication process, the differences in DS and RP values in consecutive stages are calculated and compared with the set threshold for judgment. The verification method adopted is to define indicators such as the false alarm rate, missed detection rate, and authentication rate, and use the MATLAB satellite communication toolbox\cite{MATLAB_SatCom_Toolbox}.
Nesrine Benchoubane et al.\cite{benchoubane2024securing} proposed a location-based authentication framework. Optical laser ranging technology is used to accurately measure the distance. The actual distance is compared with the planned distance, and distance verification is carried out in combination with a time-varying error threshold. If the absolute difference is within the threshold and meets the mission and orbital dynamics requirements, the connection is considered potentially secure; otherwise, the connection is rejected. The verification method adopted is to evaluate the performance of the framework through Monte Carlo simulation\cite{MonteCarlo_calculator}.
Zhifei Wang et al.\cite{wang2022random } proposed a two-step random access (TRAS) and seamless MC-controlled handover (SMHOS) method for 5G-satellite networks. The random access preamble format is redesigned to simplify the access process, reduce access delay, and improve access success rate. The management center (MC) predicts the handover of user equipment (UE) to achieve seamless handover. The verification method adopted is to conduct MATLAB simulations and compare with traditional methods.

\subsection{ AKA Protocol Based Methods}
The basic principle of securely authenticating satellite communication system terminals based on the AKA (Authentication and Key Agreement) protocol\cite{3GPP_AKA} is to ensure the identity legitimacy between the terminal and the network through a mutual authentication and key negotiation mechanism, and generate a session key for subsequent communication encryption. The AKA protocol is widely used in mobile communication systems (such as 3G/4G/5G), and its core idea can be adapted to satellite communication systems. The authentication process of this type of method is as follows: A legitimate terminal first registers in the authentication server, and is assigned a unique identifier (IMSI) and a long-term key K. The authentication server uses the terminal's unique identifier and long-term key to generate an authentication vector AV = {RAND, AUTN, XRES, CK, IK}, and then sends the authentication vector AV to the satellite communication system. The satellite network sends the random number (RAND) and authentication token (AUTN) in AV to the terminal for a security challenge. After receiving RAND and AUTN, the terminal uses the assigned long-term key K to generate an expected response RES, and then sends it to the satellite network for verification. The satellite network compares the received RES from the terminal with XRES in AV. If they are consistent, the security authentication is passed. After the authentication is passed, the terminal conducts encrypted communication with the server via the satellite.
Khan et al.\cite{khan2025modified } proposed an improved AKMA framework for decentralized continuous authentication in LEO satellite-Internet of Things (IoT) networks. Through local key refreshing for seed generation, update, and refresh, a customized transmission mode for IoT devices is achieved. It specifically includes an initial time-slot selection, a time-slot update mechanism, and an AKMA key update mechanism. A non-linear feedback shift register (NFSRs) is used to generate the transmission mode. The satellite and IoT devices independently generate seeds and transmission modes, and identity authentication is performed by comparing the transmission modes. The verification method adopted is to evaluate the performance of the authentication framework through simulation and emulation experiments.
Cheng Gong et al.\cite{gong2023distributed } proposed a distributed authentication protocol applicable to 5G satellite converged networks in the literature "A Distributed Authentication Protocol for 5G Satellite Converged Network", which combines the 5G AKA authentication scheme and the distributed system consensus scheme. When a user registers, the subscription permanent identifier (SUPI) is hidden as a subscription concealed identifier (SUCI) and sent to the home network (HN) for verification. In the access authentication stage, the UE sends an authentication request to the satellite, and the authentication is completed through the interaction among entities such as the satellite, gateway station, AUSF, and UDM. The verification method adopted is to conduct simulation evaluation.

\subsection{Physical Hardware Based Methods}
The basic principle of terminal security authentication in satellite communication systems based on physical hardware is to utilize the hardware characteristics of terminal devices (such as unique identifiers, physical unclonable functions (PUFs), security chips, etc.) as authentication factors to ensure the uniqueness and legitimacy of terminals\cite{talgat2024enhancing}\cite{abdelsalam2025physical}. The authentication process of this type of method is as follows: The terminal obtains or generates a unique hardware identifier (such as a PUF response, serial number, etc.), generates a public-private key pair based on the hardware characteristics, signs the authentication information using the private key, and sends an authentication request to the authentication server. The authentication server stores the hardware identifiers and public keys of all legitimate terminals and verifies the signature using the terminal's public key. If they match, the security authentication is passed. After successful authentication, the terminal conducts encrypted communication with the server via satellite\cite{sedjelmaci2023enabling}.
X. Ren et al.\cite{ren2023novel} proposed a lightweight access authentication scheme based on PUF, which includes an unmanned aerial vehicle (UAV) access authentication protocol and a ground terminal access authentication protocol. The PUF technology is used to achieve device identity recognition and key generation. This method does not require key storage and can resist physical attacks. At the same time, a fast handover authentication protocol for terminals when switching UAVs was proposed. The verification methods adopted include using multiple security analysis tools such as Scyther, AVISPA, and BAN-Logic for formal verification.
Igor A. Kalmylov et al.\cite{kalmykov2023application} proposed a satellite authentication protocol based on modular residue class code (MRCC) in the literature "Application of Modular Residue Classes Codes in an Authentication Protocol for Satellite Internet Systems". The MRCC is used to convert large-integer operations into parallel operations on small moduli. Based on zero-knowledge proof (ZKP) technology, satellite identity authentication is achieved in the identification system, effectively reducing the authentication time. The verification method adopted is to conduct circuit modeling tests on a Kintex UltraScale FPGA and compare the execution time of the satellite authentication protocols
Liwei Xu et al.\cite{xu2023sslpuf} proposed an access authentication and key distribution scheme for the space-air-ground integrated network based on semiconductor superlattice physical unclonable function (SSL-PUF) in the literature "An SSL-PUF Based Access Authentication and Key Distribution Scheme for the Space-Air-Ground Integrated Network". In the system model, terminals are equipped with ordinary SSL chips, and satellites and ground servers are equipped with matching SSL pairs. The authentication process is divided into three stages: terminal registration, authentication, and handover authentication. In the terminal registration stage, CRP is generated and stored. In the authentication stage, access authentication and key distribution between satellites and terminals are achieved based on the matching SSL pairs and pre-stored CRP. In the handover authentication stage, the continuity of network services when the terminal switches satellites is guaranteed. The verification methods adopted include informal analysis and using Mao Boyd Logic\cite{MaoBoyd_original} for formal security analysis.

\section{Comparison of Security Authentication Methods}
This section conducts a comprehensive comparative analysis of the above five types of satellite communication system security authentication methods from four dimensions: security, implementation difficulty and cost, applicable scenarios, and real-time performance.

\subsection{Security}
The advantage of the cryptography-based method is that it encrypts communication data through encryption algorithms to ensure the confidentiality, integrity, and authenticity of the data. Digital signature technology can verify the source and integrity of messages, preventing messages from being tampered with and forged. The key management system can ensure the secure distribution and update of keys, enhancing the security of the system. The disadvantage of this method is that key management is complex, and there is a risk of key leakage. With the emergence of quantum computing, some traditional encryption algorithms may face the threat of being cracked.

The advantage of the Blockchain-based method is that the distributed ledger technology of the Blockchain ensures the immutability and traceability of data. The decentralized feature avoids the risks of single-point failures and attacks on centralized institutions. Smart Contracts can automatically execute authentication rules, reducing human intervention and fraud risks. The disadvantage of this method is that the transaction confirmation delay is relatively high, which may not be suitable for satellite communication scenarios with high real-time requirements. The storage and computing resource requirements of the Blockchain are large, and satellite nodes may be difficult to fully meet.

The advantage of the satellite orbit observation-based method is that it uses the uniqueness and observability of satellite orbits for authentication, with high physical security. Orbital parameters are difficult to forge or tamper with, providing a reliable basis for identity authentication. The disadvantage of this method is that it requires high-precision orbit observation equipment and technology, with high costs. It is vulnerable to interference from the space environment, affecting the observation accuracy and authentication accuracy. It also requires a high degree of satellite autonomy to deal with possible observation errors and abnormal situations.

The advantage of the AKA authentication protocol-based method is that the AKA authentication protocol ensures the authenticity of the identities of both communication parties through a two-way authentication mechanism. It can effectively prevent man-in-the-middle attacks and identity spoofing attacks, ensuring the security of communication. The disadvantage of this method is that it relies on the centralized authentication server of the mobile operator, with a single-point failure risk. The execution process of the protocol may be affected by network delays and attacks, resulting in authentication failures or delays\cite{wang2022efficient }.

The advantage of the physical hardware-based method is that it realizes authentication by embedding hardware devices, such as smart cards and hardware tokens, with high security and anti-tampering properties. Hardware devices can store keys and authentication information, preventing software-level attacks. The disadvantage of this method is that users are bound to hardware devices, with poor flexibility. The production, distribution, and management costs of hardware devices are high, and they need to be closely integrated with the satellite communication system.

\subsection{Implementation Difficulty and Cost}
The implementation of the cryptography-based method requires selecting appropriate encryption algorithms and key management schemes and implementing them in satellite communication devices and ground terminals. Factors such as algorithm complexity, key length, and key update mechanisms need to be considered to ensure a balance between security and performance. At the same time, it also needs to be integrated with existing communication protocols, which may involve some compatibility issues. The implementation cost is mainly software development cost. The verification method adopted by this method is to use formal verification tools such as Scyther, AVISPA, or BAN-Logic, conduct non-formal security analysis from multiple security attributes (such as mutual authentication, key negotiation, and resistance to various attacks), and compare with existing schemes in terms of computational cost, communication cost, energy cost, and storage cost. Both formal and non-formal verification tools are open-source, with low learning difficulty and easy to master\cite{wazid2023authentication }.

The implementation of the Blockchain-based method requires building a Blockchain network architecture in the satellite communication system, including node deployment, consensus mechanism selection, and Smart Contract writing. It is necessary to solve the problems of limited computing power, storage capacity, and communication bandwidth of satellite nodes, while ensuring the security and performance of the Blockchain and preventing problems such as forks and double-spending. In addition to software development costs, the implementation cost also requires consuming a large amount of computing resources (Gas fees) to maintain the Blockchain network. The verification method adopted is to build a prototype system using Ethereum or conduct experiments on the NS2 network simulation platform, conduct non-formal security analysis from multiple security attributes, and compare with existing schemes in terms of computational cost, communication cost, energy cost, and storage cost. Both Ethereum and the NS2 simulation tool are open-source, with low learning difficulty and easy to master.

The implementation of the satellite orbit observation-based method requires high-precision orbit observation equipment and technology, as well as complex orbit calculation and analysis algorithms. It is necessary to ensure the accuracy and real-time nature of the observation data, consider the impact of space environment factors on the orbit, such as atmospheric drag and changes in the Earth's gravitational field. In addition, a satellite orbit database needs to be established and effectively integrated with the authentication system. The implementation cost is high, as it is necessary to purchase and maintain high-precision orbit observation equipment, such as radars and optical telescopes. To ensure the accuracy and real-time nature of the observation, multiple observation stations may need to work together, increasing the infrastructure construction and operation costs. The verification method adopted is to build a simulation experiment system using Matlab or the STK satellite simulation toolkit, and observe satellites at different altitudes in the Fixed-Satellite Service (FSS) and Mobile- Satellite Service (MSS) scenarios. Both Matlab and the STK satellite simulation tools are commercial software, very expensive, and have a high learning difficulty, making them difficult to master.

The implementation of the AKA authentication protocol-based method requires implementing access control in the satellite communication system and integrating it with the AKA authentication center of the operator to ensure the reliability and efficiency of the protocol in the satellite network environment. It is necessary to consider the impact of factors such as satellite link delay and packet loss on the authentication process and make corresponding optimizations. The implementation cost lies in the construction and maintenance of the authentication server and the development and integration of the AKA protocol module in communication devices. The verification method adopted is to conduct a simulation evaluation by building a LEO satellite network scenario in OPNET to verify the performance of the protocol in access authentication and handover authentication scenarios, including indicators such as response time and delay. The OPNET simulation tool is commercial software, and it is necessary to be very familiar with the AKA underlying protocol, with a high learning difficulty and difficult to master.

The implementation of the physical hardware-based method requires designing and producing special physical hardware devices, such as smart cards and hardware tokens, and ensuring their compatibility and security with satellite communication devices. The design of hardware devices needs to consider characteristics such as anti-attack and anti-tampering. At the same time, corresponding interfaces and communication protocols need to be designed for interaction with the satellite communication system. The implementation cost includes high research and development, production, and distribution costs of hardware devices, especially for some hardware devices with high-security requirements, special materials and processes are required. The verification method adopted is to use FPGA to implement a hardware prototype system for verification, and formal and non-formal tools can also be used for security verification. The hardware verification method based on FPGA cannot be mastered in the short term, and the formal and non-formal methods used are similar to other methods.

\subsection{Applicable Scenarios}
The cryptography-based method is applicable to any satellite communication system. This method authenticates users logging into the satellite communication system through user passwords, digital certificates, etc., ensuring that only authorized users can access system resources. It is suitable for various scenarios that require user identity recognition. This method can also authenticate satellite communication terminal devices. By storing keys or certificates in terminal devices, encrypted communication and identity verification are carried out with the network side to prevent illegal terminals from accessing the network and ensure the security of the communication system. This method can also be used for communication authentication between satellites and between satellites and ground stations. Through encryption keys and digital signature technology, the authenticity of the identities of both communication parties and the integrity of data are ensured, which is suitable for scenarios such as satellite networking communication and communication between satellites and ground control centers.

The Blockchain-based method is applicable to any satellite communication system. This method uses the distributed ledger and Smart Contract technology of the Blockchain. User identity information can be securely recorded on the Blockchain to achieve decentralized user identity authentication. For a large number of distributed satellite communication terminal devices, the Blockchain can provide a decentralized authentication method to ensure the authenticity and legality of the identities of terminal devices, preventing terminal devices from being tampered with or counterfeited. It is suitable for the authentication management of a large number of terminal devices in the satellite Internet of Things. This method can also be used for identity authentication and trust establishment between satellite nodes. By recording the identity information and communication history of satellite nodes in the distributed ledger, secure communication and collaboration between satellites are achieved, which is suitable for scenarios such as satellite constellation networking and secure authentication of inter- satellite communication.

The satellite orbit observation-based method is applicable to any satellite communication system. This method cannot directly authenticate users because users usually do not have the ability and conditions for satellite orbit observation. This method can be used to authenticate satellite communication terminals with specific orbital parameters. This method is mainly used for the identity authentication and status monitoring of satellites themselves. By accurately observing the orbital parameters of satellites and comparing them with the preset orbital information, the identity of the satellite and whether there are any abnormalities can be confirmed. It is suitable for scenarios such as the initial authentication after satellite launch into orbit, the identity verification during satellite in-orbit operation, and security monitoring.

The AKA authentication protocol-based method is applicable to satellite-4G/5G convergent networks because it needs to rely on the operator's authentication center. This method can achieve user identity verification and authorization through the interaction between users and the authentication server. This method can authenticate satellite communication terminal devices to ensure a secure connection between the terminal devices and the network and prevent illegal terminals from accessing the network. It is applicable to various satellite communication terminals, including fixed terminals and mobile terminals. The AKA authentication protocol can be used to authenticate the identity of satellites to ensure that ground stations communicate with legitimate satellites, ensuring the security and reliability of the satellite communication system.

The physical hardware-based method requires specific hardware support and is therefore applicable to satellite communication systems with hardware upgrades. This method cannot directly authenticate users. This method integrates physical hardware into satellite communication terminal devices. For example, by installing an encryption chip in the terminal device, the terminal device is authenticated through the unique identifier and encryption function of the hardware device. This method can equip satellites with special physical hardware devices for satellite identity authentication and security protection, such as encryption modules and identity recognition chips on satellites. Through the special functions of hardware devices, the communication security and identity authenticity of satellites are guaranteed.

\subsection{Real-Time Performance}
The cryptography-based authentication method has high real-time performance and can complete data encryption and decryption in a short time, making it suitable for satellite communication scenarios with high real-time requirements. For asymmetric encryption algorithms, although their computational complexity is high, with the support of hardware acceleration technology, they can still meet the real-time requirements of most satellite communication applications. However, when it comes to the encryption processing of a large amount of data or complex authentication protocols, a certain delay may be introduced. But through reasonable algorithm design and optimization, real-time performance can still be ensured within an acceptable range.

The Blockchain-based method usually has lower real-time performance than the cryptography-based method due to the distributed nature of Blockchain technology and the consensus mechanism, which makes its authentication process relatively complex. In a Blockchain network, each node needs to verify the legality of transactions and reach a consensus, which involves multiple steps and a certain time delay. For example, in public Blockchains, consensus mechanisms such as Proof-of-Work (PoW)\cite{POW_security} are used, and nodes need to perform a large amount of calculations to verify the legality of blocks, resulting in a long confirmation time, which is not suitable for satellite communication scenarios with extremely high real-time requirements. Although private Blockchains and consortium Blockchains can improve the transaction confirmation speed by optimizing the consensus mechanism, they still cannot compare with the fast authentication methods based on cryptography.

The real-time performance of the satellite orbit observation-based method depends on the accuracy of the observation equipment and the data processing speed. Generally, it takes a certain amount of time to obtain and analyze satellite orbit data, especially when high-precision orbit information is required. Multiple observations and complex data processing algorithms may be needed. For example, during the initial orbit determination stage after satellite launch, it is necessary to observe the satellite multiple times through ground observation stations or other satellites and calculate using orbital mechanics models. This process may take several hours or even several days to obtain accurate orbit information, with poor real-time performance. Currently, with the use of some high-precision radar observations and fast data processing algorithms, the orbital parameters of satellites can be obtained and authenticated within a few minutes to tens of minutes.

The AKA authentication protocol-based method is designed to meet the fast identity authentication requirements in mobile communications and has good real-time performance. It realizes user identity authentication and key negotiation through fast message interaction between the user terminal and the network side. In satellite communication, the AKA authentication protocol can efficiently transmit authentication messages using the channel resources of the satellite network, and the design of the protocol itself is concise and efficient, enabling the authentication process to be completed in a short time.

The physical hardware-based method usually has high real-time performance. Physical hardware such as encryption chips and smart cards integrates special encryption algorithms and processing circuits, which can quickly complete operations such as encryption, decryption, and identity authentication. Using an encryption chip can encrypt and decrypt data in real-time when data is sent and received, with almost no obvious delay. Moreover, the authentication process of physical hardware is relatively simple, mainly through the unique identifier of the hardware device and the pre-stored key for verification, without the need for complex calculations and network interactions, so it can quickly complete authentication and is suitable for satellite communication scenarios with extremely high real-time requirements.

\subsection{Comparison Results}
Based on the above multi-dimensional comparative analysis, the comparison results of satellite communication system security authentication methods shown in Table 1 can be obtained. The most suitable and optimal satellite communication system security authentication method should be the Blockchain based method, with a score of 17 points; followed by the cryptographic based method, with a score of 16 points; then the physical hardware based method, with a score of 14 points; the AKA protocol based method is relatively poor, with a score of 9 points; and the method with the lowest score and the worst performance is the satellite orbit information based method, with only 8 points.

\begin{table*}[htbp]
\centering
\caption{Comparison Results of Security Authentication Methods for SatCom}
\label{tab:security_methods}
\begin{tabular}{|l|c|c|c|c|}
\hline
\textbf{Method} & \textbf{Security} & \textbf{Implementation Difficulty \& Cost} & \textbf{Applicable Scenarios} & \textbf{Real-Time Performance} \\ \hline
Cryptographic-based methods & Medium & Good & Best & Good \\ \hline
Blockchain-based methods & Best & Good & Best & Medium \\ \hline
Satellite orbit-based methods & Best & Worst & Worst & Worst \\ \hline
AKA protocol-based methods & Medium & Poor & Poor & Poor \\ \hline
Physical hardware-based methods & Best & Poor & Poor & Best \\ \hline
\end{tabular}
\footnotesize
\\[1ex]
\textbf{Note:} Best = 5 points, Good = 4 points, Medium = 3 points, Poor = 2 points, Worst = 1 points.
\end{table*}

\section{Prospects for Future Research Directions}
In the future, the field of security authentication in satellite communication systems has broad research space and development directions. In terms of new technology applications, the development of quantum communication technology has brought new opportunities to security authentication in satellite communication systems. Quantum communication has the property of unconditional security. Based on the principles of quantum mechanics, it utilizes the non-clonability of quantum states and the phenomenon of quantum entanglement to achieve absolutely secure information transmission. In satellite communication systems, studying how to combine quantum communication technology with existing security authentication methods and developing security authentication protocols based on quantum key distribution will be an important research direction in the future\cite{fu2022ztei }.

In order to effectively resist the threat of password cracking brought about by future quantum computing, a new generation of cryptographic security authentication methods resistant to quantum computing will also be an important research direction in the future. For example, Wang et al.\cite{wang2022ntru} proposed a lightweight certificateless anonymous access authentication scheme based on NTRU (N-th Degree Truncated Polynomial Ring Units). The NTRU key generation algorithm was improved for system key-pair generation to resist quantum attacks and save satellite resources; a certificateless anonymous communication algorithm was designed to achieve entity anonymous authentication through identity encryption and hashing, reducing satellite-ground interactions. The security of NTRU depends on the Shortest Vector Problem (SVP)\cite{SVP_original} and the Closest Vector Problem (CVP)\cite{CVP_vanEmdeBoas} in the NTRU lattice theory. These problems are also considered difficult to solve under quantum computing and can resist the attacks of future quantum computers.

Dharminder et al.\cite{dharminder2023post} proposed a post-quantum secure authentication protocol based on Ring-Learning with Errors (RLWE). The protocol includes setup, registration, login and authentication, and password change phases. It uses technologies such as bio-hashing and random numbers to achieve identity authentication and key negotiation through multiple interactions between users and the Network Control Center (NCA) server, effectively resisting quantum computing attacks. The verification methods adopted are two ways: informal security analysis and formal security analysis. Informal analysis demonstrates the security of the protocol from multiple aspects such as user anonymity, integrity, and resistance to man-in-the-middle attacks; formal analysis is based on the random oracle model, and by defining a series of games to simulate the behavior of attackers, it proves the security of the protocol under certain conditions. At the same time, by comparing with other protocols, the performance advantages of this protocol are shown29\cite{li2021effective }.

With the deep integration of satellite communication systems with technologies such as the Internet of Things and artificial intelligence, it is also crucial to study security authentication technologies suitable for integrated scenarios. In the satellite Internet of Things, a large number of Internet of Things devices transmit data via satellites. How to ensure the identity authentication of these devices and the security of data transmission is a key problem to be solved. By combining Blockchain technology, a distributed identity authentication system for Internet of Things devices can be constructed. Utilizing the immutability and traceability of Blockchain, the authenticity of device identities and the integrity of data can be ensured.

In the scenario of the integration of artificial intelligence and satellite communication systems, studying how to use artificial intelligence technology to conduct real-time analysis and processing of security authentication data, and achieve intelligent early warning and automatic response to security threats is also one of the future research focuses. For example, P. Fu et al.\cite{peng2025zero} proposed a security authentication method for satellite communication systems based on the Zero Trust Architecture (ZTA)\cite{NIST_ZTA} and edge intelligence. This method extends the traditional zero-trust concept to multiple dimensions, covering subjects, objects, environments, behaviors, and physical entities. Based on this architecture, a continuous authentication scheme is designed. Through periodic monitoring and re-evaluation of variable attributes, active continuous authentication is achieved, and an edge intelligence algorithm based on Neural-Based Decision Trees (NBDTs)\cite{NBDT_original} is used to improve the authentication accuracy. The verification method adopted by this method is to implement a prototype system using Python and conduct verification from three aspects: security protection performance, startup evaluation, and runtime evaluation. By simulating requests from different sources, the authentication accuracy of different schemes for static and dynamic requests is compared; the response time of the system under different startup conditions is evaluated; and the runtime performance of continuous authentication under different load conditions is tested.

In terms of improving the security authentication system, further strengthening the research and formulation of security authentication standards for satellite communication systems and promoting the unification and coordination of international standards will help improve the security and interoperability of global satellite communication systems. With the globalization of satellite communication systems, satellite communication system systems in different countries and regions need to be interconnected. Unified security authentication standards can ensure the secure docking between different systems. Studying how to establish a more perfect security authentication supervision mechanism, strengthening the supervision of satellite communication system service providers and equipment manufacturers to ensure the effective implementation of security authentication measures, is also a direction that needs to be paid attention to in the future. Establishing a security authentication evaluation index system to quantitatively evaluate the performance and security of satellite communication system security authentication systems provides a basis for the optimization of the security authentication system \cite{oligeri2022pastai }.

In response to new security challenges, with the continuous development of satellite communication systems, new security threats will continue to emerge. Studying how to prevent the impact of space debris on satellite communication system security authentication systems and how to deal with new types of network attack means, such as quantum attacks and artificial-intelligence-driven attacks, will be important research topics in the future. As satellite communication systems are increasingly widely used in key fields such as military, finance, and energy, studying how to ensure the security authentication of satellite communication systems in key fields and ensure the security of national critical information infrastructure also has important practical significance\cite{liu2021decentralized }. By establishing a multi-level and multi-dimensional security protection system, strengthening the security monitoring and emergency response of satellite communication systems in key fields, the anti-attack ability and recovery ability of satellite communication systems in key fields can be improved.

\section{Conclusion}
By means of literature research and comparative analysis, this paper carries out on a comprehensive survey towards security authentication methods used by SatCom. This paper first summarizes the existing security authentication methods for SatCom as five categories, i.e., cryptography based methods, Blockchain based methods, satellite orbit information based methods, AKA protocol based methods and physical hardware based methods. Then, it reviews the authentication processes and verification methods used by each category. Subsequently, a comprehensive comparative analysis is carried out on the above-mentioned five types of security authentication methods from four dimensions, i.e., security, implementation difficulty and cost, applicable scenarios and real-time performance, and the final comparison results are obtained. Finally, prospects are made for several important future research directions of security authentication methods for SatCom, laying a well foundation for further carrying on the related research works.


\newpage
\begin{IEEEbiography}[{\includegraphics[width=1in,height=1.25in,clip,keepaspectratio]{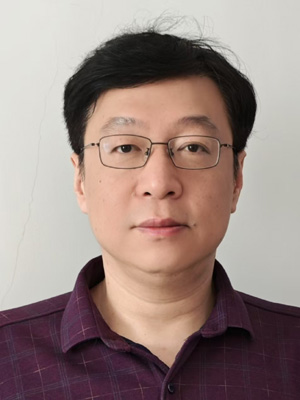}}]{Yunfei Meng}
received a Bachelor of Engineering degree in Computer Science and Technology from Shandong University, China in 2001. In 2004, he obtained a Master of Engineering degree in Software Engineering from the same university. In 2021, he was awarded a Doctor of Engineering degree in Software Engineering from Nanjing University of Aeronautics and Astronautics, China. Since 2022, he has been serving as Assistant Professor at the College of Information Engineering, Qingdao Binhai University, China. His research interests mainly include the security authentication methods, security model based on zero-trust architecture, SDN/NFV techniques as well as formal methods. He has published several influential papers in some important academic journals such as IEEE TNSM, Elsevier C\&S, Elsevier FGCS and etc..
\end{IEEEbiography}

\begin{IEEEbiography}[{\includegraphics[width=1in,height=1.25in,clip,keepaspectratio]{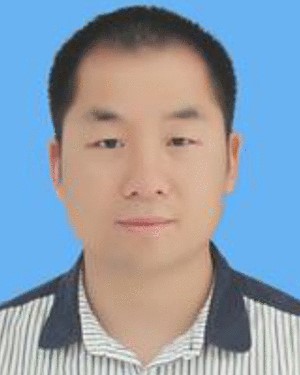}}]{Changbo Ke}
received a Bachelor of Engineering degree in Computer Science and Technology from Kunming University of Science and Technology, China, in 2008. In 2010, he obtained a Master of Engineering degree from the same university. In 2014, he was awarded a Doctor of Engineering degree in Computer Science and Technology from Nanjing University of Aeronautics and Astronautics, China. Currently, he serves as Associate Professor at the Schoole of Computer Science, Nanjing University of Posts and Telecommunications, China. His research interests mainly include the privacy preservation for IoT systems and formal methods. He has published several papers in some important academic journals, such as IEEE TSC, IEEE IOT, Elsevier KBS and etc..
\end{IEEEbiography}

\begin{IEEEbiography}[{\includegraphics[width=1in,height=1.25in,clip,keepaspectratio]{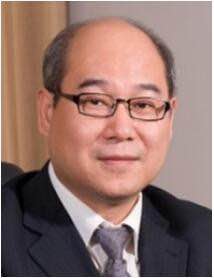}}]{Zhiqiu Huang}
received a Bachelor of Engineering degree in Computer Science and Technology from National University of Defense Technology, China in 1987. In 1990, he obtained a Master of Engineering degree in the same major from the same university. In 1999, he was awarded a Doctor of Engineering degree from Nanjing University of Aeronautics and Astronautics, China. Currently, he serves as Full Professor and doctoral supervisor at the College of Computer Science and Technology, Nanjing University of Aeronautics and Astronautics, and also serves as the director of Key Laboratory for Safety Critical Software Development and Verification (Nanjing University of Aeronautics and Astronautics), Ministry of Industry and Information Technology. His research interests mainly include the safety-critical software systems, privacy preservation and formal methods.
\end{IEEEbiography}

\vfill

\end{document}